\documentclass[journal=jpccck,manuscript=article]{achemso}
\usepackage{amsfonts}
\usepackage{amsmath}
\usepackage{amssymb}
\usepackage{graphicx}
\usepackage[usenames,dvipsnames]{xcolor}

\newcommand{\mc}[1]{\mathcal{#1}}

\author{Ra\'{u}l A. Bustos-Mar\'{u}n}
\email{rbustos@famaf.unc.edu.ar}
\affiliation[IFEG and FAMAF, UNC]
{IFEG (CONICET) and Facultad de Matem\'{a}tica Astronom\'{i}a y F\'{i}sica
, Universidad Nacional de C\'{o}rdoba, ,5000, C\'{o}rdoba, Argentina}

\author{Eduardo A. Coronado}
\affiliation[INFIQC and FCQ, UNC]
{INFIQC (CONICET) and Facultad de Ciencias Qu\'{i}micas, Universidad Nacional de C\'{o}rdoba,
5000, C\'{o}rdoba, Argentina}

\author{Horacio M. Pastawski}
\affiliation[IFEG and FAMAF, UNC]
{IFEG (CONICET) and Facultad de Matem\'{a}tica Astronom\'{i}a y F\'{i}sica
, Universidad Nacional de C\'{o}rdoba, ,5000, C\'{o}rdoba, Argentina}

\title{Excitation-Transfer Plasmonic Nanosensors based on Dynamical Phase Transitions.}
\begin{document}

\begin{abstract}
Dynamical Phase transitions (DPTs) describe the abrupt change in the dynamical properties of
open systems when a single control parameter is slightly modified.
Recently we found that this phenomenon is also present in a simple model of a linear array
of metallic nanoparticles (NPs) in the form of a localized-delocalized DPT.
In this work we show how to take advantage of DPTs in order to design a new kind of plasmonic
sensor which should own some unique characteristics.
For example, if it were used as plasmon ruler it would not follow the so called universal
plasmon ruler equation [\textit{Nano Letters} \textbf{2007}, 7, 2080-2088], exhibiting instead
an \textit{on-off} switching feature.
This basically means that a signal should only be observed when the control/measured parameter,
\textit{i.e.} a distance in the case of plasmon rulers, has a very precise and pre-determined value.
Here, we demonstrate their feasibility and unique characteristics, showing that they combine high
sensitivity with this \textit{on-off} switching feature in terms of different distances
and local dielectric constants. This property has the potentiality to be used in the design of new
plasmonic devices such as plasmonic circuits activated only under certain environmental conditions.

\end{abstract}
\section{INTRODUCTION}

The study of localized surface plasmon (LSP) resonances over the last couple of decades
has led to great advances in several areas of science and technology.
Probably, its most significant application is its use in
nanosensors.\cite{CRNuzzo,CRHafner,CRNordlander,Coronado1,Coronado2,Coronado3,
Coronado-Ruler,Rong,El-Sayed1,El-Sayed2,El-Sayed3,Ben-Park,Cherqui,Funston,Huang,Liu-Li,
FRET1,FRET2,FRET3,FRET4,FRET5,FRET6,SERS1,SERS2,SERS3,SERS4}
Most of the strategies used to turn a plasmonic device into a sensor are based on the
shift of the resonant position of the LSP\cite{CRNuzzo,CRHafner,CRNordlander,
Coronado1,Coronado2,Coronado3,Coronado-Ruler,Rong,El-Sayed1,El-Sayed2,El-Sayed3,
Ben-Park,Cherqui,Funston,Huang,Liu-Li} or in exploiting hot spots to
increase spectroscopic signals.\cite{SERS1,SERS2,SERS3,SERS4}
In this respect, there is a growing interest in using interacting plasmonic systems to measure
distances at the nanoscale. These devices are known as nanorulers and they are finding
interesting applications in several fields such as biology.\cite{Rong}
Most of the proposed plasmonic rulers work by the same principle, the shift of a resonance
induced by the variation of a distance.
As noticed originally by Jain et. al.,\cite{El-Sayed1} the frequency shift of the LSP resonance
follows an almost universal exponential or quasi-exponential law with respect to the separation
between NPs.
This behavior is known as the universal plasmon ruler equation (UPRE) and was
observed in simulations and experiments using NPs of different size and
shape,\cite{El-Sayed1,El-Sayed2,El-Sayed3} core-shell NPs,\cite{Liu-Li} and even two dimensional
arrays of NPs.\cite{Ben-Park}
There are some deviations of the UPRE due to retardation effects or the effects of higher order
multipoles, but they can usually be corrected by empirical fittings.\cite{Coronado-Ruler}
Only when NPs are almost touching each other, there are some deviations, nonfitable to
exponential, of the UPRE.\cite{Funston,Huang}
There is another kind of plasmon rulers that is now being actively investigated. They are based
on a different principle, the fluorescence resonance energy transfer (FRET). Its working
principle is the resonant transference of excitation from a fluorescent molecule to a
nearby NP. The closer to fluorophore to the NP, the greater the photoluminescence quenching.
This effect has been successfully used as a nanorulers in several experimental
examples.\cite{FRET1,FRET2,FRET3,FRET4,FRET5,FRET6}

Collective effects induced by LSP couplings in NPs arrays can give rise to other phenomena
and new potential applications.\cite{libros1,libros2,Ejem-CDA1,Ejem-CDA2,Ejem-CDA3,Ejem-CDA4,Ejem-CDA5,Ejem-NoCDA,Nature-2003,CDA1,CDA2,CDA3,CDA4,CDA5,CDA6}
One example is the use of Fano resonances to design plasmonic sensors.\cite{CRNordlander}
In this case, the shift of very sharp peaks or valleys in the extinction spectrum, consequence
of collective effects of interacting NPs, have been proposed to measure distances or the
dielectric constant of the surrounding medium.\cite{CRNordlander}
More exotic applications of NPs coupling are also
possible and many of them are now being actively explored, such as clocking by
metamaterials or plasmonics circuitry.\cite{Invisibility,Circuitry}

\begin{figure}[ht]
\begin{center}
\includegraphics[width=2.5in]{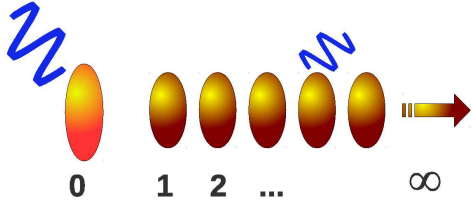}
\end{center}
\caption{Scheme of the system studied. An external source locally excites
the LSP of a NP which is coupled to a waveguide where detection takes place.
The arrow on the right indicates that the array is infinite in this direction.}
\label{Figure-1}
\end{figure}

Although, linear arrays of metallic NPs have been proposed as potential waveguides to transfer
massive amounts of information at the nanoscale,\cite{Brongersma} nowadays it is clear that
high damping factors would impose severe limits to that and most of the effort in this field
is being put on overcoming this limitation.
However, one important and very general feature of 1-D systems, which seems to have been
overlooked in plasmonics, is the extreme sensitivity of their dynamical properties to slight
changes of their parameters.
In this respect, we have previously studied the plasmonics energy transfer from a locally
excited NP (LE-NP), the 0th NP in the scheme shown in \ref{Figure-1},
to the interior of a semi-infinite NP chain.\cite{Bustos}
This system presents a form of dynamical phase transition (DPT), a phenomenon that is
currently attracting interest as source of novel effects on various
fields.\cite{PT-breaking1,PT-breaking2,Dente,Rotter1,Rotter2}
Basically a DPT in this context means that by sightly moving a single
parameter, such as the shape of the NPs or the distance between them, the system undergoes an
abrupt transition from transferring all the injected excitation to the interior of the NP array
(a delocalized state), to keeping all the external excitation in the NPs closer to the excitation
point (a localized state).\cite{Bustos} The aim of this work is to
exploit this basic property into a realistic setting, in order to design a new kind of
excitation-transfer plasmonic nanosensor (ETPS).
\section{THEORY}

The system studied, depicted in \ref{Figure-1}, is modeled using the coupled dipole
approximation.
\cite{Brongersma,Bustos,CDA1,CDA2,CDA3,CDA4,CDA5,CDA6,Multipoles,Ejem-CDA1,Ejem-CDA2,Ejem-CDA3,Ejem-CDA4,Ejem-CDA5}
In this model, each $i^{\mathrm{th}}$-NP is described by a dipole $P_{i}$ induced by the
electric field produced by the other dipoles, $E_{i,j}$, and the external source,
$E_{i}^{(\mathrm{ext})}$.
We assume a generic ellipsoidal shape for the metallic NPs and describe its polarizabilities
in a quasi-static approximation,\cite{Ellipsoids} resulting in:
\begin{equation}
P_{i}=\frac{\epsilon _{0}\mathcal{V}_{i}(\epsilon_{i} -\epsilon _{m,i})}{\left[
\epsilon _{m,i}+L_{i}(\epsilon_{i} -\epsilon _{m,i})\right] } \left( E_{i}^{(%
\mathrm{ext})}+\sum\limits_{j\neq i}E_{i,j} \right) \label{Pvector},
\end{equation}%
where $\mc{V}$ is the volume, $L$ is a geometrical factor that depends on the
shape of the NP and the direction of $E$, $\epsilon _{0}$ is the free
space permittivity, and $\epsilon_{m,i}$ is the dielectric
constant of the host medium around the $i$-th NP.
The dielectric constants $\epsilon_{i}$ of the NPs are
described by a Drude-Sommerfeld model: $\epsilon = \epsilon_{\infty}-\frac{\omega _{%
\mathrm{p}}^{2}}{(\omega ^{2}+i\omega \eta )}$, where $\omega _{\mathrm{p}%
}^{}$ is the plasmon frequency, $\eta$ is the electronic damping factor, and
$\epsilon_{\infty}$ is a material dependent constant that take into account
the contribution of the bound electrons to the polarizability.
The LSP oscillations over each NP can only be transverse ($T$) or
longitudinal ($L$) to the axis of the linear array and we assume, for
simplicity, that the ellipsoids axes are aligned with respect to these $T$ and $L$
directions. If the excitation wavelength is large compared with the
separation between NPs, $E_{i,j}$ can be taken in the near field approximation, $E_{i,j}=%
\frac{-\gamma ^{T,L}P_{j}}{4\pi \epsilon _{0}\epsilon
_{m}d_{i,j}^{3}}$, where $\gamma ^{T}=1$, $\gamma ^{L}=-2$ and $d$ is the
distance between NPs. Taking into account these considerations, $P_{i}$ and $E_{i}^{(%
\mathrm{ext})}$ can be arranged as vectors $\mathbf{P}$ and $\mathbf{E}$
resulting in:\cite{Bustos}
\begin{equation}
\mathbf{P}=\left( \mathbb{I}\omega ^{2}-\mathbb{M}\right) ^{-1}\mathbb{R}%
\mathbf{E},  \label{MatrixP}
\end{equation}%
where $\mathbb{M}$ is the dynamical matrix and $\mathbb{R}$ is a diagonal matrix that
rescales the external applied field according to local properties:
\begin{equation}
R_{i,i}=-\epsilon _{0} \mathcal{V}_{i} \omega_{_{\mathrm{P}i}}^{2}f_i, \label{Rii}
\end{equation}%
with $f_i=\frac{ \left[ 1 -
(\epsilon_{\infty}-\epsilon _{m,i}) \left( \omega ^{2}-i\omega \eta_{i}\right)
/\omega_{_{\mathrm{P}i}}^{2} \right] }
{\left[\epsilon _{m,i}+L_{i}(\epsilon_{\infty}-\epsilon _{m,i})\right]}$.

Because of the cubic dependence of $E$ on $d$, we neglect contributions beyond nearest
neighbors,\cite{Brongersma, Bustos}
resulting in a tridiagonal $\mathbb{M}$.  For NPs belonging to
the chain, $(i,j)\neq 0,$ $M_{i,i}=\omega _{_\mathrm{SP}}^{2}-\mathrm{i}\eta \omega$ and
$M_{i,j}=\omega _{_\mathrm{X}}^{2}$, and for the LE-NP,\ $M_{0,0}=\omega _{_{\mathrm{SP}0}}^{2}%
-\mathrm{i}\eta_{0}\omega $, $M_{0,1}= \omega _{_{\mathrm{X}0,1}}^{2}$ and
$M_{1,0}=\omega _{_{\mathrm{X}1,0}}^{2}$. The LSP frequencies $\omega _{_{\mathrm{SP}i}}^{}$
and coupling constants $\omega _{_{\mathrm{X}ij}}^{2}$, result:
\begin{equation}
\omega _{_{\mathrm{SP}i}}^{2} =\frac{\omega _{_{\mathrm{P}i}}^{2}L_{i}}{\left[
\epsilon _{m,i}+L_{i}(\epsilon_{\infty}-\epsilon _{m,i})\right] },
\label{OmegaSP}
\end{equation}
\begin{equation}
\omega _{_{\mathrm{X}ij}}^{2} =\frac{\gamma ^{_{T,L}} \mc{V}_{i} \omega _{_{\mathrm{P}i}}^{2}}
{4\pi \epsilon_{m}d_{ij}^{3}} f_i.
\label{OmegaX}
\end{equation}
For spherical NPs ($L=1/3$) in vacuum ($\epsilon _{m}=1$) within a Drude-Sommerfeld model
which completely neglects contribution from bound electrons ($\epsilon _{\infty}=1$),
Eqs. \ref{MatrixP}-\ref{OmegaX} reduce to the simple form used in several
references.\cite{Bustos, Brongersma,CDA1,CDA2,CDA3,CDA4,CDA5,CDA6}
This shows the meaning of $f_i$, which essentially collects the deviations from
the ideal model of spherical NP in vacuum with $\epsilon_{\infty}=1$, not only for $R_{i,i}$
but also for couplings between NPs. Note that these deviations reduce to a constant for
$\epsilon_{\infty} \approx \epsilon_{m,i}$ and/or small excitation frequencies $f_i$ (strictly for
$[\omega^2-i \omega \eta_i] \ll \omega_{_{\mathrm{P}i}}^{2}/ [\epsilon_{\infty}-\epsilon_{m,i}]$).
The form of Eqs. \ref{MatrixP}-\ref{OmegaX} not only extends our previous expressions\cite{Bustos}
to non-spherical NPs but also includes explicitly the interdependencies of
$\omega _{_{\mathrm{X}ij}}^{2}$, $\omega _{_{\mathrm{SP}i}}^{2}$, and $R_{i,i}$ with the
parameters of the system, \textit{i.e.} size, shape, and material of NPs, distances between NPs,
and dielectric constants.
It should also be mentioned that although this model can be further improved, the different corrections
will only add quantitative corrections as long as the essential physics of the problem remains, \textit{i.e.}
a semi-infinite chain with nearest neighbors interactions.\cite{Bustos, Dente, GF}
We will analyze this point in more detail in the last section.

In the present work we will consider the case of excitation injection to a semi-infinte
homogeneous linear array of NPs, where solely the LE-NP can be different from the rest and
the only nonidentical separation is that between the LE-NP and the first NP of the chain.
Therefore, there will be two LSP resonances, that of the LE-NP $\omega_{_{\mathrm{SP}0}}^{}$ and
that of the chain's NPs $\omega_{_\mathrm{SP}}^{}$. Similarly, the couplings between the
chain's NPs will be all the same $\omega_{_\mathrm{X}}^{2}$. The couplings between the LE-NP
and the first NP of the chain are $\omega_{_{\mathrm{X}0,1}}^{2}$ and
$\omega_{_{\mathrm{X}1,0}}^2$. They are not necessarily equal (see Eq. \ref{OmegaX}),
since they describe quite different situations, \textit{i.e.} how $P_1$ affects $P_0$
($\omega_{_{\mathrm{X}0,1}}^2$) and how $P_0$ affects $P_1$ ($\omega_{_{\mathrm{X}1,0}}^2$).

The values of the resonance frequency $\omega_{_\mathrm{SP}}^{}$ and the couplings
$\omega_{_{\mathrm{X}ij}}^{2}$ can be tuned in different ways. However, one must be careful
because they are always interrelated.
For example, choosing a different material for the NPs will change both $\omega_{_P}$ and
$\epsilon_\infty$ and therefore $\omega_{_{\mathrm{X}ij}}^{2}$ and $\omega_{_\mathrm{SP}}^{}$.
Changing  the    shape of the  NPs  will modify $L$ and consequently
$\omega_{_{\mathrm{X}ij}}^{2}$ and $\omega_{_\mathrm{SP}}^{}$,
except for $\epsilon_\infty=\epsilon_m$ in which case it will only modify
$\omega_{_\mathrm{SP}}^{}$. The volume and distances between NPs will change
$\omega_{_{\mathrm{X}ij}}^{2}$. If the material, size and shape of NPs are
fixed, the free parameters that can be used for sensing purposes are the dielectric constant
of the medium $\epsilon_m$ that alter both $\omega_{_{\mathrm{SP}i}}^{}$ and
$\omega_{_{\mathrm{X}ij}}^{2}$, and the distances between NPs that change couplings.

The values of $\omega_{_\mathrm{X}}^{2}$ and $\omega_{_\mathrm{SP}}^{}$ define the passband
which is the interval of frequencies at which an excitation can propagate with only a relatively
small decay, given by $\eta$, through the waveguide formed by the linear array of NPs.
All excitations with frequencies $\omega$ outside the passband will decay along the chain
exponentially and very fast.
In the weak damping limit (WDL) $\eta \rightarrow 0$, this passband is given by
a simple expression,
$\frac{|\omega^2-\omega_{_\mathrm{SP}}^2|}{\omega_X^2}\leq 2$.
This fact shows one of the roles of the chain: to determine which frequencies can
be propagated. The other role is to perturb the local density of plasmonic
states (LDPS) of the LE-NP by pushing the resonance from its uncoupled value at
$\omega_{_{\mathrm{SP}0}}^{}$ to the nearest edge of the passband.\cite{Bustos,Dente,GF}
This process occurs very abruptly when the resonance is about to cross the passband edge
that is when the dynamics of the system changes completely.\cite{Bustos,Dente}

This phenomenon can be completely understood in terms of the divergences, or poles, of Eq.
\ref{MatrixP} in the WDL.\cite{Bustos} Basically there are three different dynamical regimes
of interest for excitation transfer: 1) \textit{delocalized} or \textit{resonant}
state regime, where the maximum of LDPS of the LE-NP
falls within the passband and most of the excitation can be transferred to the NP array;
2) \textit{localized} state regime where there is a very sharp peak in the LDPS outside the
passband and most of the excitation remains close to the excitation point;
and 3) a transition regime called \textit{virtual} state where the LDPS
presents a non-Lorentzian asymmetric peak just at the passband edge and most of the excitation
can be transferred to the waveguide.
The transitions between the different dynamical regimes
can be obtained analytically in the WDL and assuming
frequency independent couplings, where:\cite{Bustos}
\begin{equation}
\alpha =\frac{2\beta +4 V^2-\beta ^{2}-1}{4 V^2},  \label{alfaC1}
\end{equation}
gives the \textit{resonant-virtual} transition and:
\begin{equation}
\alpha =2\pm \frac{(1-\beta )}{V},  \label{alfaC2}
\end{equation}
gives the \textit{virtual-localized} transition, where $\alpha = (\omega _{%
_{\mathrm{X}1,0}}^{2}\omega_{_{\mathrm{X}0,1}}^{2}/\omega _{_\mathrm{X}}^{4})$,
$\beta =(\omega _{_{\mathrm{SP}0}}^{2}/\omega _{_\mathrm{SP}}^{2})$, and
$V=\left( \omega _{_\mathrm{X}}^{2}/\omega _{_\mathrm{SP}}^{2}\right)$. The constant
$\alpha$ is the relative coupling between LE-NP and first NP of
the chain in units of the coupling in the chain, $\beta$ yields the difference
in resonance frequency of the LE-NP with respect to that of a chain's NP,
and $V$ measures the strength of the coupling between chain's NPs
in units of their own resonance frequency.
These formulas, although strictly valid only in the WDL, give an excellent
estimation for finites $\eta$s.\cite{Bustos}

The excitation transferred from the LE-NP to the \textit{m}-th NP of the
chain can be calculated exactly (in the quasi-static limit and within the nearest-neighbors
approximation) by using the formula:\cite{Bustos}
\begin{equation}
P_{m} =
\frac{\omega_{_{\mathrm{X}1,0}}^{2} R_{00} E_{0}}
{\omega _{_\mathrm{X}}^{2} \left( \left[ \omega ^{2}-\widetilde{\omega }_{\mathrm{SP}0}^{2}\right]
-\alpha \Pi (\omega ) \right )} e^{-m (l \pm iq)}, \label{Pm}
\end{equation}
where $E_{0}$ is the electric field on the LE-NP and the renormalized eigen-frequency
$\widetilde{\omega }_{_{\mathrm{SP}}}$ is given by $\widetilde{\omega }_{_{\mathrm{SP}}}^{2}=%
\omega_{_{\mathrm{SP}}}^{2}-i \eta \omega$.
The decay length $l$ and the wavenumber $q$ depend on the self energy $\Pi$ as:
\begin{equation}
l \pm iq = \ln (\omega_{_{\mathrm{X}}}^{2}/\Pi),
\end{equation}
where $\Pi$ is:
\begin{align}
\Pi (\omega )& =\tfrac{1}{2}\left[ \omega ^{2}-\widetilde{\omega }_{_{\mathrm{%
SP}}}^{2}\right] -  \label{Pi} \\
& \mathrm{sgn}(\omega ^{2}-\omega _{_{\mathrm{SP}}}^{2})\tfrac{1}{2}\sqrt{\left[
\omega ^{2}-\widetilde{\omega }_{_{\mathrm{SP}}}^{2}\right] ^{2}-4\omega _{_{%
\mathrm{X}}}^{4}}.  \notag
\end{align}

In all the plots of the next section with use $|P_{m}|^2$ as an
indicator of the excitation transferred to the waveguide, with
$\omega _{_\mathrm{X}}^{2}/\omega _{_\mathrm{SP}}^{2}=0.45$, $\eta /\omega _{_\mathrm{SP}}^{}=0.03$,
and using $m=8$. We have chosen $m=8$, because around this value the behavior of $|P_{m}|^2$ vs $m$
becomes essentially a small decaying exponential for all $\omega$ within the passband or zero for
excitation frequencies outside it. Thus, choosing a bigger value of $m$ would only add a multiplicative
factor to the results, \textit{i.e.} it would not change qualitatively the figures shown.
\section{RESULTS}
\begin{figure}[ht]
\begin{center}
\includegraphics[width=3.5in]{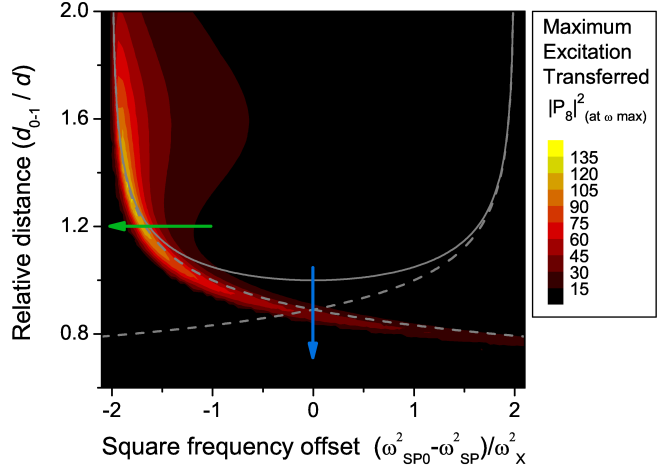}%
\end{center}
\caption{Maximum value, allowed by a variation of the excitation frequency,
of the excitation transferred to a NP of the waveguide as function of the system's parameters.
gray lines indicate DPTs: the continuous line corresponds to the
\textit{resonant-virtual} DPT and the dashed lines show the \textit{virtual-localized}
transitions.}
\label{Figure-2}
\end{figure}

In this work, our purpose is to show how to use the sudden change in the dynamical properties
of systems tuned around their DPT to device ETPSs. For this purpose, we used the system
depicted in \ref{Figure-1} as an example. \ref{Figure-2} shows the maximum excitation
transferred to a NP of the waveguide, $|P_{8}|^{2},$ enabled by a variation of the excitation
frequency $\omega$, at each system configuration. Superposed are the critical values that
separates dynamically distinct behaviors calculated in the WDL (Eqs. \ref{alfaC1} and \ref{alfaC2}).
As it has been previously reported,\cite{Bustos} excitation transfer is controlled by
the DPT where \textit{virtual} states are transformed into \textit{localized} states
(gray dashed lines).

\begin{figure}[ht]
$\begin{array}{c}
\includegraphics[width=2.5in]{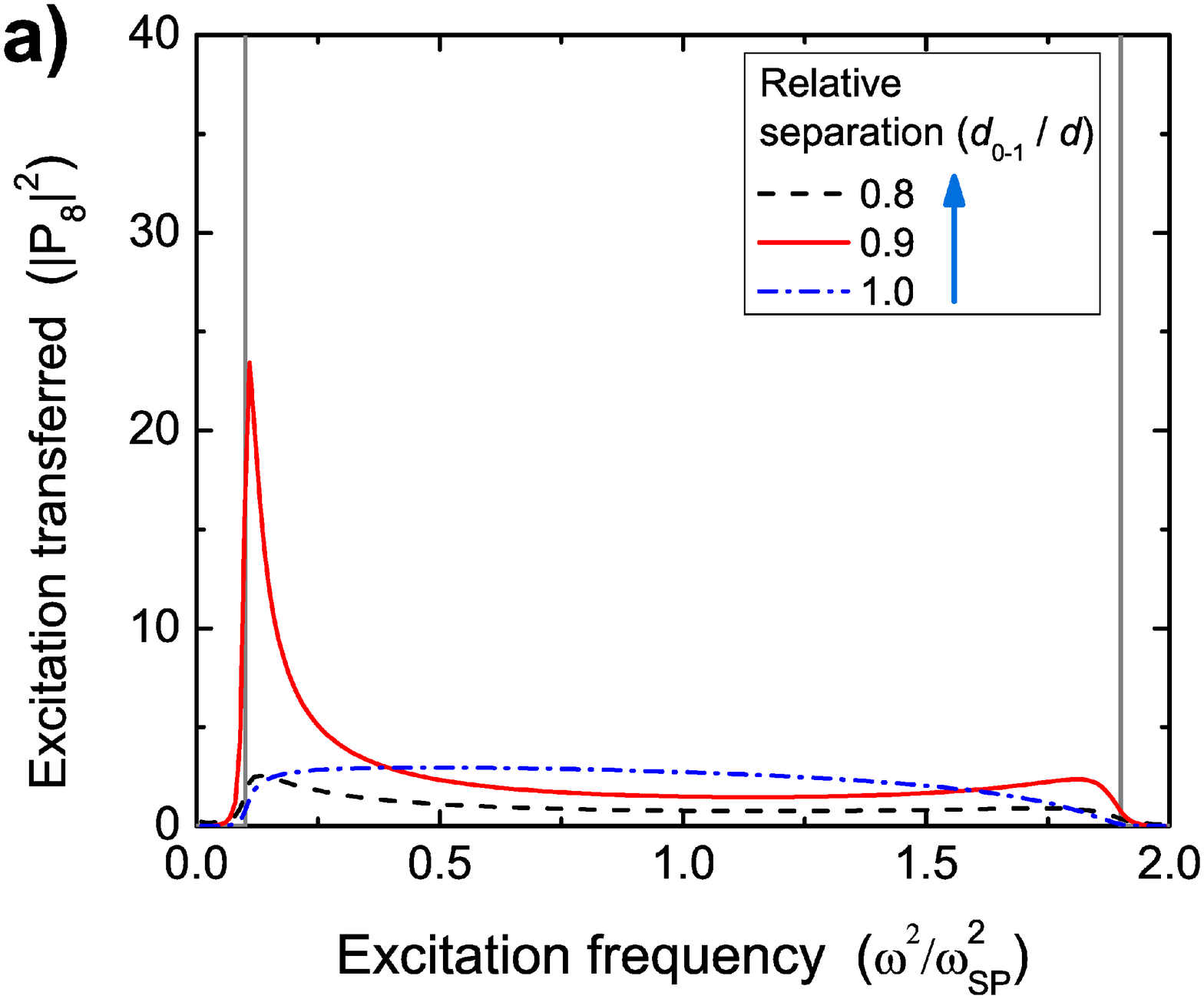} \\
\includegraphics[width=2.5in]{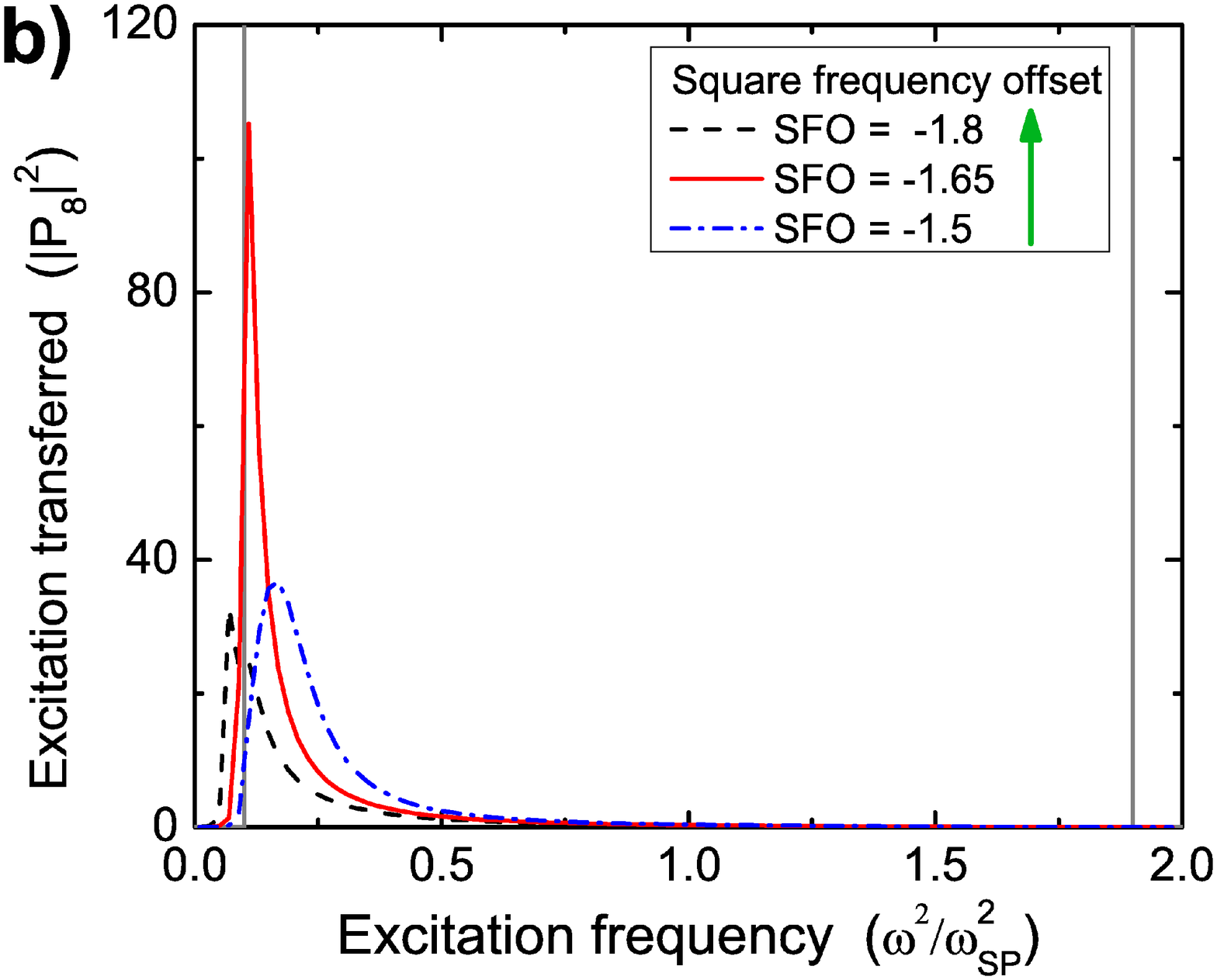}%
\end{array}$
\caption{\textbf{a)} Excitation transferred to the 8th NP of the waveguide,
$|P_8|^2$ in arbitrary units, as a function of $\omega$ for three different values of $d_{0-1}/d$.
Each spectrum corresponds to a point along the vertical blue arrow shown in \ref{Figure-2}.
The value of $\lvert P_{8}\lvert^{2}_{\mathrm{(at \ \omega \ max)}}$ in \ref{Figure-2}
corresponds here to the maximum value of $\lvert P_{8} \lvert^{2}$ in a given curve. The two
vertical grey lines mark the passband edges of the waveguide in the WDL.
\textbf{b)} The same but moving the SFO for fixed $(d_{0-1}/d)$ along the horizontal
green arrow shown in \ref{Figure-2}.}
\label{Figure-3}
\end{figure}

Analyzing the phase diagram of \ref{Figure-2} it is clear that the
strong dependence of excitation transfer to the system's parameters can be
used for different forms of sensors. As a first application of this,
lets us analyze the potentiality of this plasmonic device as a plasmon ruler.
In \ref{Figure-2} a vertical displacement corresponds to a variation of the distance between
the LE-NP and the chain.
The vertical blue arrow indicates an example of a region in the parameters
space that can be used for such a purpose. For each point along this
arrow the system depicts a different excitation transfer spectrum.
Three representative points along this line, at $d_{0-1}/d=0.8$, $0.9$, and $1.0$,
are used in \ref{Figure-3}-\textbf{a}) to illustrate how the spectrum of excitation
transfer changes when the system undergoes a DPT. Note the abrupt change in the spectra
induced by a slight change of $d_{01}/d$: the excitation transfer suddenly increases at
$\omega^2/\omega^2_{_\mathrm{SP}} \cong 0.1 $ when $d_{01}/d$ varies from $1.0$ to $0.9$,
and then decreases equally abruptly when $d_{01}/d$ change from $0.9$ to $0.8$.
This behavior is a consequence of the \textit{virtual-localized} DPT, indicated
in \ref{Figure-2} by the curved gray dashed lines and
predicted by Eqs. \ref{alfaC1} and \ref{alfaC2}.
One of the interesting aspects of this new form of plasmonic ruler
is that the variation of a distance not merely modifies the position of a peak
but dramatically changes the whole spectrum, and quite remarkably,
before and after the DPT, only negligible signals should be
observed independently of the excitation frequency. In this sense, our excitation
transfer nanoruler could be used as a robust nanometric standard.
This is so, because, in order to observe a difference in the excitation transferred,
it is not required a high accuracy in the excitation frequency nor
a monochromatic light but keeping
the maximum of the excitation frequency  close
to the lower band edge frequency, in our example at $\omega^2/\omega_{_\mathrm{SP}}^2=0.1$,
is a sufficient condition.
Furthermore, if  the polarization has two components one transversal, $T$, and one
longitudinal ${L}$, this should not interfere in a sensor tuned to work with only one of them.
This is so, because the passband of $T$ modes are half of $L$ modes, Eq. \ref{OmegaX}, which imply that
regions in the spectrum where DPTs can be observed, should be well separated.
Moreover, according to Eqs. \ref{alfaC1} and \ref{alfaC2},
DPTs should occur for different values of system's parameters.
Another important aspect is that the detection could be performed in a region encompassing several NPs.
This is so, because, except for the first NPs of the chain, the excitation transferred
to the chain is an exponential function of NP position $m$, see Eq. \ref{Pm}.
Thus, the main feature of this kind of nanorulers is its robust \textit{on} and \textit{off}
feature, \textit{i.e.} signals should only be detected when the LE-NP is at a very precise and
predetermined distance.
\begin{figure}[ht]
\begin{center}
$\begin{array}{c}
\includegraphics[width=2.5in]{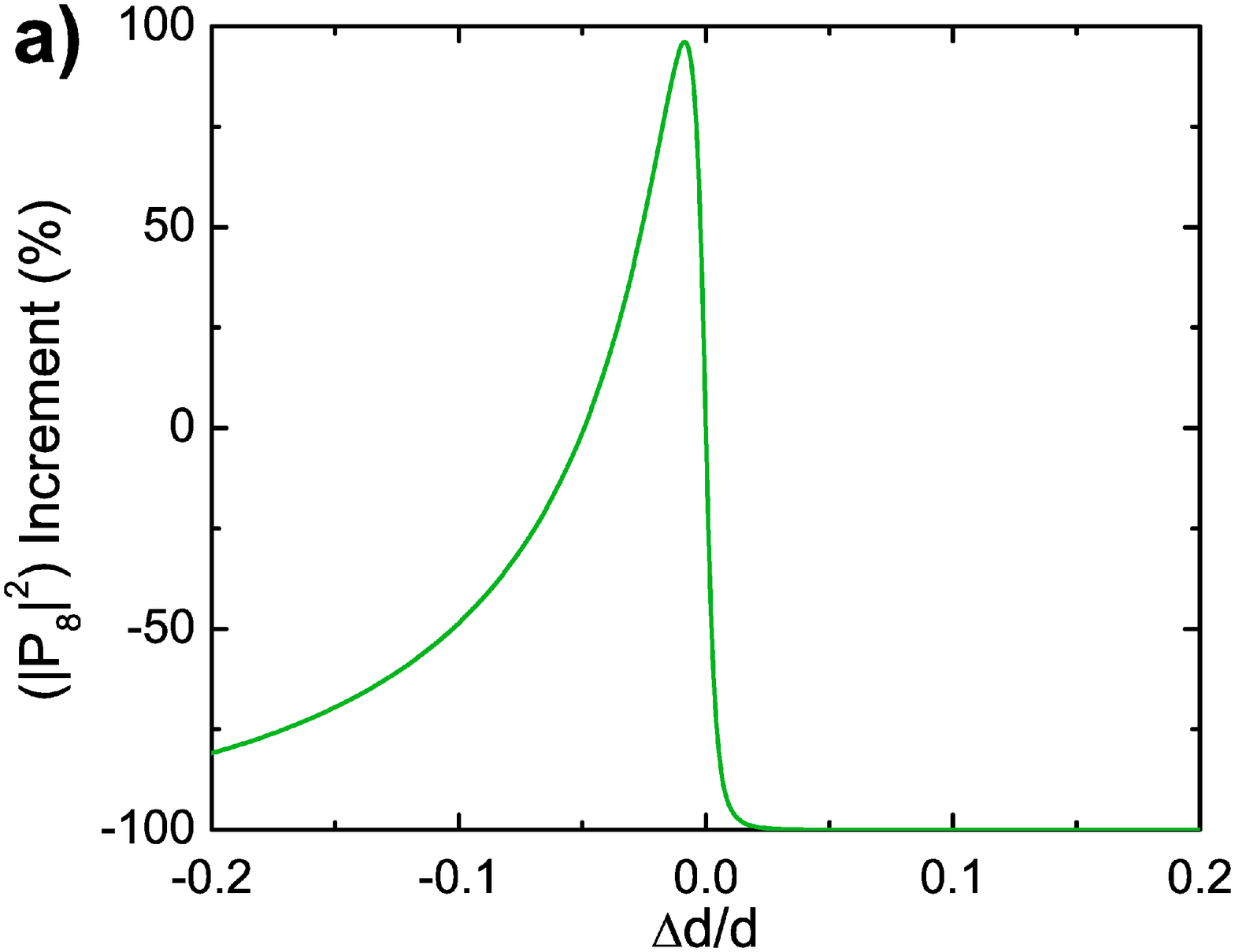} \\
\includegraphics[width=2.5in]{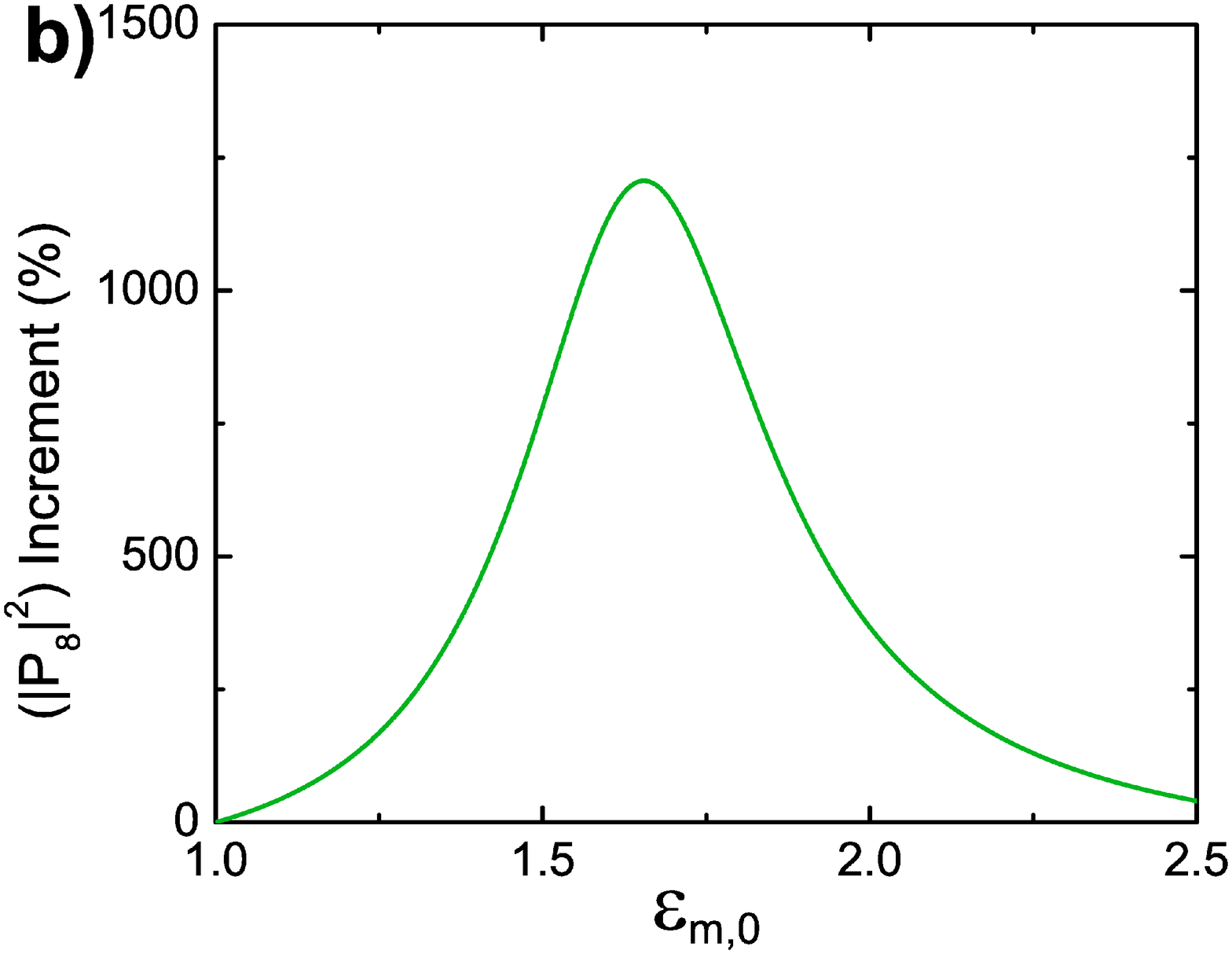}%
\end{array}$
\end{center}
\caption{\textbf{a)} Relative increment of the excitation transferred,
$(|P|^{2}-|P|^{2}_{initial})/|P|^{2}_{initial}$, as a function of the relative expansion of
the whole system, $\Delta d/d_{initial}$, for $\omega^2/\omega_{_\mathrm{SP}}^2=0.1$.
The initial conditions are $d_{0,1}/d=0.97$ and SFO=-1.58.
\textbf{b)} The same but changing the local dielectric constant around the LE-NP,
$\epsilon_{m,0}$. The initial conditions are $d_{0,1}/d=0.83$ and SFO=-1.33.}
\label{Figure-4}
\end{figure}

ETPS could also be used for sensing other properties, not only as plasmon nanorulers.
As shown in \ref{Figure-2}, the system is also very sensitive to
the value of the square frequency offset (SFO =
$(\omega_{_{\mathrm{SP}0}}^{2}-\omega _{_\mathrm{SP}}^{2})/\omega_{_\mathrm{X}}^{2}$)
and consequently to $\omega_{_{\mathrm{SP}0}}^{2}$, $\omega _{_\mathrm{SP}}^{2}$, and/or
$\omega_{_\mathrm{X}}^{2}$.
Increasing the numerator of the SFO or decreasing its denominator, will both move
the SFO away from the center of \ref{Figure-2} as indicated by the horizontal green arrow.
\ref{Figure-3}-\textbf{b}) shows the excitation transfer spectrum for three values of SFO
along this green arrow. Similarly to the nanoruler case, the spectrum experiences a significant
change when moving the SFO along a DPT, from $SFO=-1.5$ to $-1.8$.
In order to show how to make profit of this feature we will consider two examples: one in which
contraction or expansion of the whole system is measured, and one in which
local dielectric constant around the LE-NP is measured.
\ref{Figure-4}-\textbf{a} shows, for a fixed excitation frequency,
$\omega^2/\omega_{_\mathrm{SP}}^2=0.1$,
the relative increment of the excitation transferred,
$(|P|^{2}-|P|^{2}_{initial})/|P|^{2}_{initial}$,
as a function of the relative expansion/contraction of the whole system, which
turns our device into an optical strain monitor\cite{CRNordlander} or even into
a temperature sensor, depending on the expansion coefficient of the host material.
Note that a relative expansion/contraction of the whole system corresponds
to a change in $\omega_{_\mathrm{X}}^2$ but not in $d_{0-1}/d$,
$\omega_{_{\mathrm{SP}0}}^2$, nor in $\omega_{_\mathrm{SP}}^2$.
Thus it is equivalent to an horizontal displacement in \ref{Figure-2}.
\ref{Figure-4}-\textbf{b} shows the variation of the relative increment of the excitation transferred
as a function of the local dielectric constant around the LE-NP, a property potentially
useful for molecular sensing purposes for example.\cite{CRHafner}
In this case, a variation of $\epsilon_{m,0}$ changes not only
$\omega^2_{_{\mathrm{SP}0}}$ but also $\omega^2_{_{\mathrm{X}0,1}}$ and $R_{0,0}$ (Eqs.
\ref{Rii}-\ref{OmegaX}). Therefore, it is equivalent to a diagonal displacement in \ref{Figure-2}
and a rescaling of the external applied field.
\section{RETARDATION EFFECTS AND CONTRIBUTIONS BEYOND NEAREST NEIGHBORS}

The key ingredients from which the phenomenology described so far arise are: dominant
nearest-neighbors interactions and the semi-infinite character of the system.
If this conditions are fulfilled, the results presented above should be applicable to the system under
consideration besides quantitative corrections such as shifts of the spectra or the exact position of
DPTs in the parameter's phase space. However, it is important to understand how our results are
corrected by different factors.

In this section, we will study the effect of interactions beyond nearest-neighbors and
retardation effects.
In this case, Eq. \ref{Pvector} is still valid but all coupling between NPs should be considered,
\textit{i.e.} $\mathbb{M}$ is no longer tridiagonal, and these couplings should be determined by the
dipole-induced electric field beyond the quasi-static limit, \textit{i.e.}:
\begin{equation}
\vec {E} = \frac { e^{ ik d } }{ 4\pi \epsilon d^{ 3 } } %
\left\{ (k d)^{ 2 }(\hat { d } \times \vec { p } )\times \hat { d } +%
\left[ 3\hat { d } (\hat { d } \cdot \vec { p } )-\vec { p }  \right] \left( 1- ikd \right)  \right\},
\end{equation}
where $k$ is the wavenumber in the dielectric, $k=\omega/v$ (where $v$ is the speed of light in the medium),
$\hat { d }$ is the unit vector in the direction of $\vec {d}$ (where $\vec {d}$ is the position of the
observation point with respect to the position of the dipole).

As the system consists of a linear array of NPs where the spheroids axes are aligned with respect to
the direction of the array, transversal ($T$) and longitudinal ($L$) excitations do not mix.
Thus, the coupling terms $\omega_{X}^2$ of Eq. \ref{OmegaX} acquire the form:
\begin{equation}
\omega _{_{\mathrm{X}ij}}^{2} =\frac{ \widetilde{\gamma }^{_{T,L}}_{i,j} \mc{V}_{i} \omega _{_{\mathrm{P}i}}^{2}}
{4\pi \epsilon_{m}d_{i,j}^{3}} f_i,
\end{equation}
where
\begin{align}
& \widetilde{\gamma }^{_{L}}_{i,j}=-2[1-ikd_{i,j}]e^{ ikd_{i,j} } \notag \\
& \widetilde{\gamma }^{_{T}}_{i,j}=[1-ikd_{i,j}-(kd_{i,j})^{ 2 }]e^{ ikd_{i,j} }.
\end{align}
Eq. \ref{MatrixP} has now to be solved numerically which can be done by using standard methods of
linear algebra. \cite{Numerical}

\begin{figure}[ht]
\begin{center}
\includegraphics[width=2.5in]{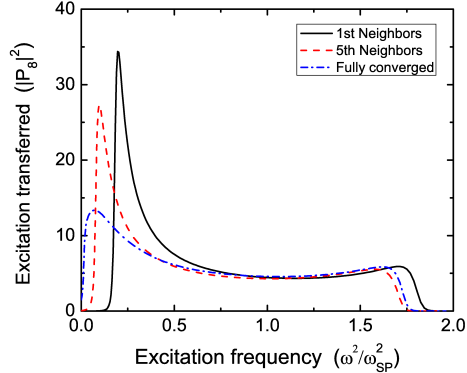}%
\end{center}
\caption{Excitation transferred to the waveguide,
$|P_8|^2$ in arbitrary units, as function of $\omega^2/\omega_{_\mathrm{SP}}^{2}$ for $d_{0-1}/d=0.9$,
$\omega _{_\mathrm{X}}^{2}/\omega _{_\mathrm{SP}}^{2}=0.41$ and $\eta /\omega _{_\mathrm{SP}}^{}=0.032$.
Different curves show numerical results within the near field approximation for: nearest-neighbor interactions,
fifth-nearest-neighbor interactions, and a fully converged calculation. The system is at the DPT for the
curve with nearest-neighbor interactions.}
\label{Figure-5}
\end{figure}

The physical system considered in this section corresponds to a linear array of very flat oblate spheroidal
NPs of silver ($\epsilon_{\infty}=3.7$, and $\omega _{\mathrm{P}}^{}=9.3$x$10^{15}$rad/s) with the radio of
the minor axis equals to $4.5$nm and the radii of the major axis equal to $46.5$nm (shape factor $L_{i}=0.86$,).
The major to minor axis ratio is about 10 which is large enough to ensure a strong quadrupole quenching.
\cite{KCS}
Separation between NPs of the chain is $13.5$nm, from center to center, and they are aligned with
the two equal axes perpendicular to the array direction.
The external electric field is applied only to the first NP where the direction of $E_{0}^{(ext)}$ is also
perpendicular to the array, \textit{i.e.} $\gamma = \widetilde{\gamma }^{_{T}}_{i,j}$.
Electronic damping is calculated using the Matthiessen's rule $\eta =v_{F}(1/l_\mathrm{bulk}+C/l_{eff})$ where
the Fermi velocity is $v_{F}=1.38$x$10^6$m/s, the bulk mean free path is $l_\mathrm{bulk}=57nm$, we use $C\cong 1$,
and the effective mean free path is calculated using $l_\mathrm{eff}=4V/S$ where $V$ is the volume and $S$
the surface of the spheroids.\cite{Eta}
For simplicity we consider a dielectric medium with $\epsilon_{m} = \epsilon_{\infty}$. Taking into account this,
$\eta /\omega _{_\mathrm{SP}}^{}$ and $\omega_{_\mathrm{X}}^{2} /\omega_{_\mathrm{SP}}^{2}$ give 0.032 and
0.41 (for $kd=0$) respectively . We use here a slightly smaller coupling than that of the previous section for
reasons that will become clear in the context of \ref{Figure-5}.
As before, the observation point is fixed at the 8th NP, and we use the value of $|P_{8}|^2$ as an indicator
of the excitation transferred to the chain.
Due to the large interparticle coupling, finite size effects result so important, especially near the lowest
band edge, that it is necessary to take very long size of chains. We use 200 NPs, a value large enough to ensure
negligible finite size effects.

\begin{figure}[ht]
\begin{center}
$\begin{array}{c}
\includegraphics[width=2.5in]{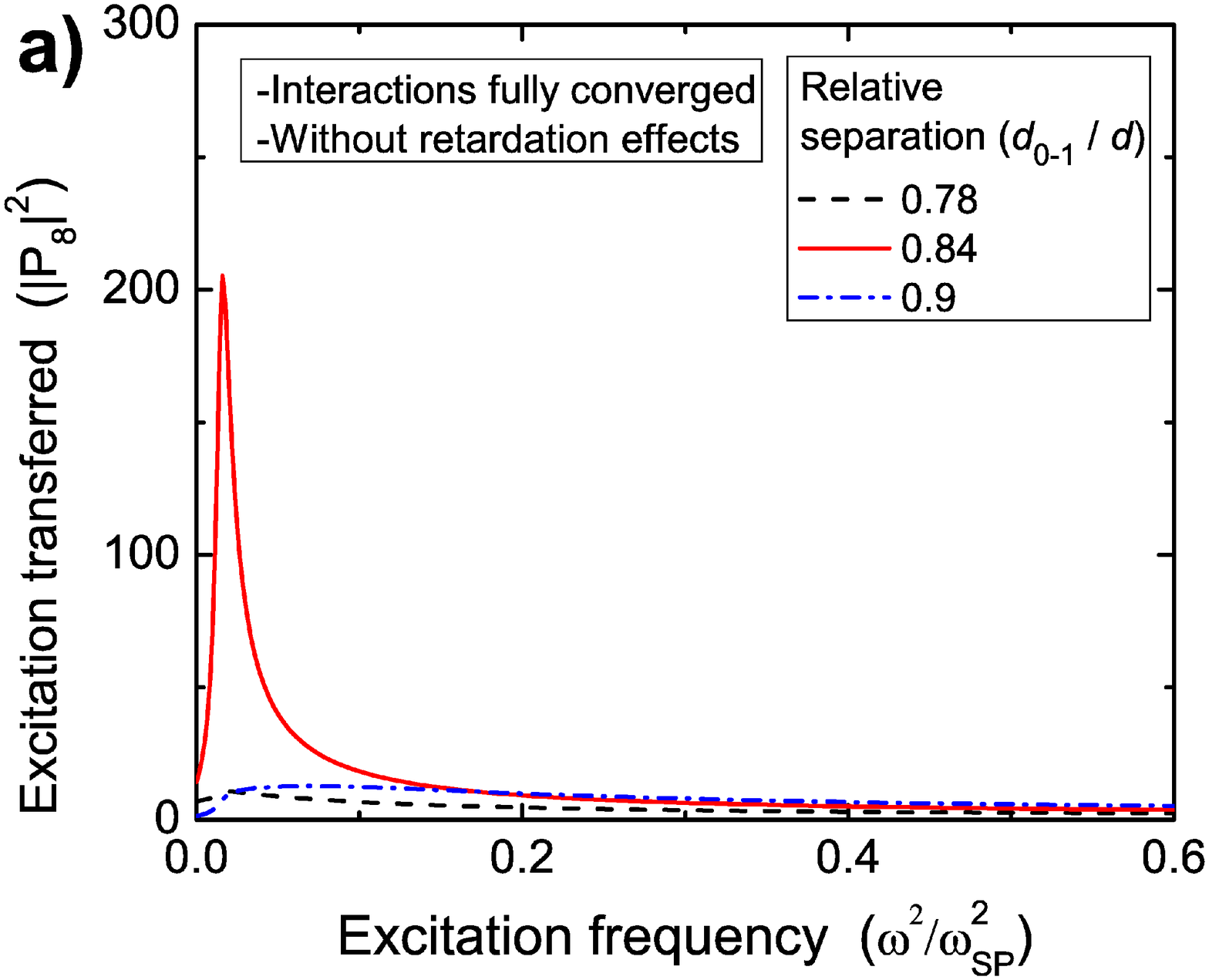} \\
\includegraphics[width=2.5in]{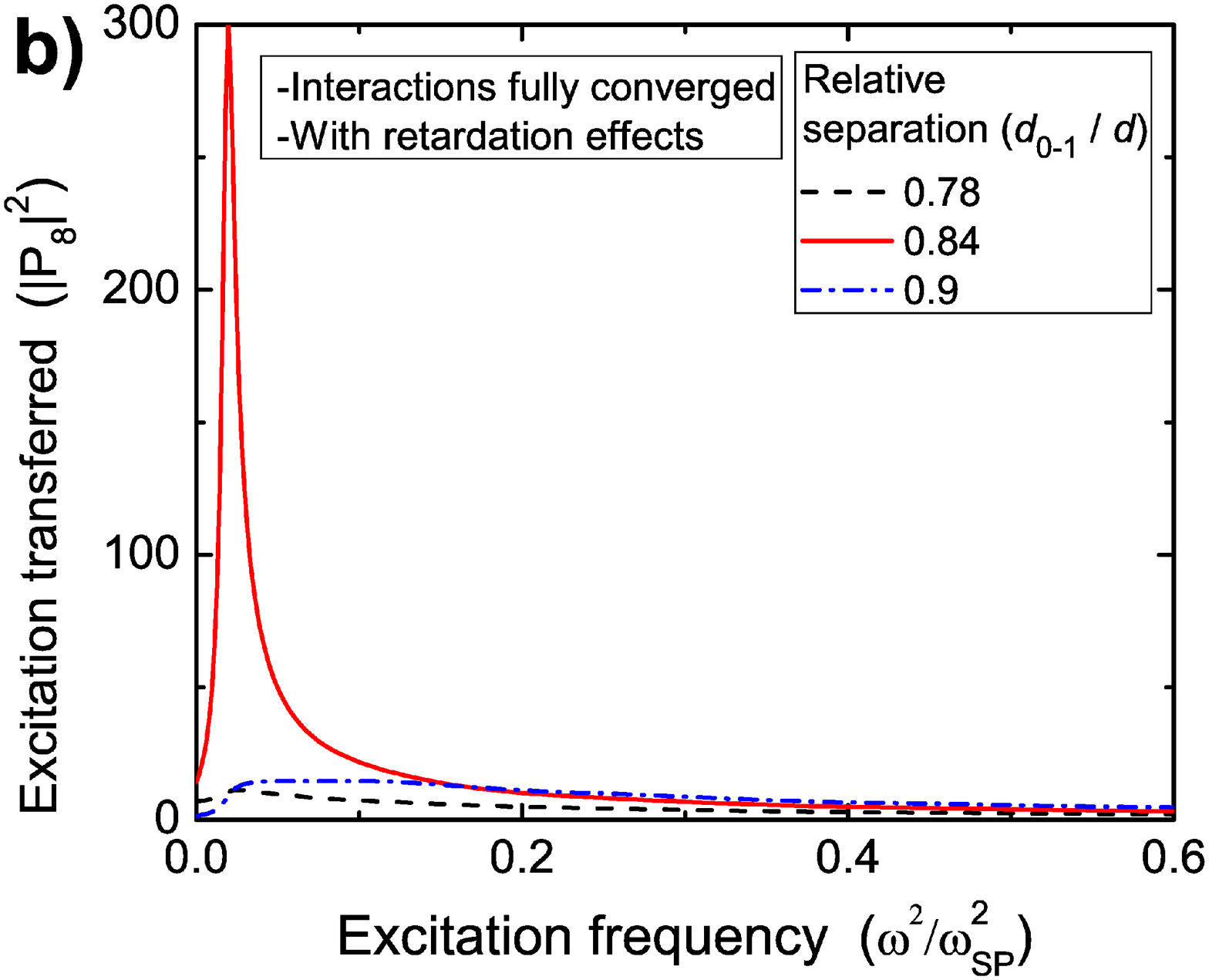}%
\end{array}$
\end{center}
\caption{\textbf{a)} Numerical results without retardation effects of the
excitation transferred to the waveguide, $|P_8|^2$ in arbitrary units, as function of
$\omega^2/\omega_{_\mathrm{SP}}^{2}$ for different values of $d_{0-1}/d$.
\textbf{b)} The same but with retardation effects.
In both cases all interactions between NPs are considered.}
\label{Figure-6}
\end{figure}

\ref{Figure-5} shows the effect of including interaction beyond nearest-neighbors. The first effect of this
is essentially a red-shift and a broadening of the passband. The second effect is that the DPT is also shifted
from $d_{0-1}/d=0.9$ to $d_{0-1}/d=0.84$ as can be seen in \ref{Figure-6}-a. However, it is still
present and indeed, as the damping term is proportional to the frequency, the peak is increased.
Note that if $\omega _{_\mathrm{X}}^{2}/\omega _{_\mathrm{SP}}^{2}=0.45$ were used, the peak would
disappear from the spectrum as it would be located at $\omega^2<0$, showing an important effect of
higher-order contributions to interactions.
In \ref{Figure-6}-a and b retardation effects on the DPT are compared. Here, retardation causes
nothing else but an increase of the intensity of the peak, in the region of the spectrum where
the DPT is observable. This is because, in this region of the spectrum,
around $\omega^2/ \omega_{_\mathrm{SP}}^2=0.02$, $d$ is much smaller than $k$ ($kd \approx 0.05$).
Of course at higher frequencies, for example at $\omega \approx \omega_{SP}$, this is not the case ($kd \approx 0.4$) and the
behavior of the system is completely beyond the near field approximation.
\begin{figure}[ht]
\begin{center}
\includegraphics[width=2.5in]{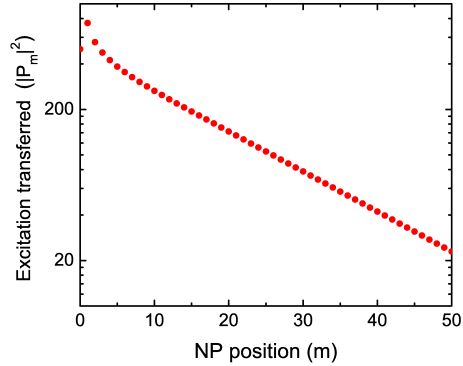}%
\end{center}
\caption{Excitation transferred to a NP of the waveguide, $|P_m|^2$ in arbitrary units, as function
of its position $m$. Retardation effects as well as all interactions between NPs are considered.
Here $d_{0-1}/d=0.84$ and the excitation frequency is $\omega^{2}/\omega _{_\mathrm{SP}}^{2}=0.02$,
which correspond to the peak of \ref{Figure-6}-b}
\label{Figure-7}
\end{figure}

Finally, \ref{Figure-7} shows the value of $|P_{m}|^2$ for different NPs of the
chain. Note that even after including all contribution to the interaction among NPs and retardation effects,
$|P_{m}|^2$ still decays exponentially as predicted by the theory, although the decay rate is different
mainly because the passband edge is considerably shifted.
\section{CONCLUSIONS}

In this work we have studied the potentiality for sensing purposes of
the excitation transferred from a LE-NP to the interior of a semi-infinite NP chain, when the
system is very close to its DPT condition.
While most plasmonic sensors use the shift of the LSP resonance peaks or the
photoluminescence quenching of hybrid plasmonic-fluorophores systems, here we introduce
a new working principle.
Basically, the idea is to take advantage of the abrupt change in the plasmonic energy transferred when
a control parameter is slightly changed around a DPT.
We have shown that this kind of sensor has the unique characteristic of having an
\textit{on-off} switching property and a high sensitivity, which opens new possibilities to
design plasmonic devices such as plasmonic circuits activated only under certain
environmental conditions.
We have also addressed the effects of different corrections to the approximations used
and have shown that if the system's parameters are chosen properly, the DPT should be observable and
therefore useful for the sensing purposes.

\begin{figure}[p21]
\begin{center}
\includegraphics[width=7.0in]{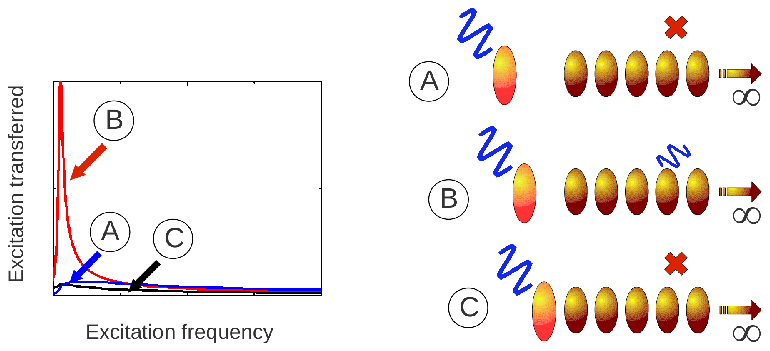}%
\end{center}
\caption{Graphical abstract.}
\label{TOC}
\end{figure}
\end{document}